\documentclass[%
preprint,
nofootinbib,
amsmath,amssymb,
aps,
prd,
showpacs,
superscriptaddress
]{revtex4-1}
\usepackage{setspace}
\usepackage[caption=false]{subfig}
\usepackage{graphicx}
\usepackage{dcolumn}
\usepackage{bm}
\usepackage{mathtools}

\DeclareMathOperator{\Tr}{Tr}

\usepackage{mathrsfs}
\usepackage{enumitem}
\usepackage[unicode=true,bookmarks=true,
bookmarksnumbered=false,bookmarksopen=false,
breaklinks=false,
pdfborder={0 0 1},
backref=false,colorlinks=true]{hyperref}

\hypersetup{pdftitle={title}, pdfauthor={ChunJun Cao, Sean M. Carroll}, citecolor=blue,linkcolor=blue,urlcolor=blue,citecolor=blue}

\usepackage{color}
\definecolor{purple}{rgb}{0.5,0,0.5}

\newcommand{\R}{\mathscr{R}}
\newcommand{\region}{\Sigma}
\newcommand{\area}{\mathcal{A}}
\newtheorem{thm}{Theorem}[section]

\newtheorem{example}[thm]{Example}

\DeclareMathOperator{\co}{:}
\begin{document}

\preprint{CALT-TH-2017-069}

\title{Bulk Entanglement Gravity without a Boundary: 
\\ Towards Finding Einstein's Equation in Hilbert Space}

\author{ChunJun Cao}
\email{ccj991@gmail.com}
\affiliation{Walter Burke Institute for Theoretical Physics, California Institute of Technology, Pasadena, CA 91125, USA}
\author{Sean M.\ Carroll}
\email{seancarroll@gmail.com}
\affiliation{Walter Burke Institute for Theoretical Physics, California Institute of Technology, Pasadena, CA 91125, USA}

\begin{abstract}
We consider the emergence from quantum entanglement of spacetime geometry in a bulk region.
For certain classes of quantum states in an appropriately factorized Hilbert space, a spatial geometry can be defined by associating areas along codimension-one surfaces with the entanglement entropy between either side.
We show how Radon transforms can be used to convert this data into a spatial metric.
Under a particular set of assumptions, the time evolution of such a state traces out a four-dimensional spacetime geometry, and we argue using a modified version of Jacobson's ``entanglement equilibrium'' that the geometry should obey Einstein's equation in the weak-field limit.
We also discuss how entanglement equilibrium is related to a generalization of the Ryu-Takayanagi formula in more general settings, and how quantum error correction can help specify the emergence map between the full quantum-gravity Hilbert space and the semiclassical limit of quantum fields propagating on a classical spacetime.
\end{abstract}

\maketitle
\tableofcontents
\vfill\eject

\baselineskip=14pt

\section{Introduction}

There has been considerable recent interest in the idea of deriving an emergent spacetime geometry from the entanglement properties of a quantum state \cite{Swingle:2009bg,VanRaamsdonk:2010pw,TNSGeo,Faulkner:2013ica,Faulkner:2017tkh,Czech:2015kbp}.
Much of this work has taken place within the context of the Anti-de Sitter/Conformal Field Theory (AdS/CFT) correspondence \cite{Maldacena:1997re}. In particular, various results related to emergent gravity  \cite{Faulkner:2013ica,Swingle:2014uza,Faulkner:2017tkh} have yielded not only confidence in the program, but also helpful insights in understanding quantum gravity itself  \cite{Maldacena:2013xja,Sanches:2016sxy}. Moreover, the idea need not be restricted to the context of holographic duality. Indeed, as it was originally proposed  \cite{VanRaamsdonk:2010pw}, it can in principle be generalized to derive other geometries closer to our own physical universe \cite{Susskind:2014moa,Cao:2016mst,Bao:2017iye,Bao:2017qmt,Susskind:2017ney}. Entanglement also plays a role in entropic/thermodynamic gravity \cite{Jacobson:1995ab,Padmanabhan:2009vy,Verlinde:2010hp,Jacobson:2015hqa,Carroll:2016lku,Verlinde:2016toy} and the holographic-spacetime approaches of Banks and Fishler \cite{Banks:2010tj,Banks:2011av,Banks:2015iya} and of Nomura et al. \cite{Nomura:2016aww,Nomura:2016ikr}.

The real world, needless to say, does not seem to have anti-de~Sitter boundary conditions.
It is therefore interesting to ask whether we can derive spacetime and gravitational field equations directly in the bulk from the entanglement properties of quantum states. 
The emergence of a semiclassical spacetime description from a quantum state, thought of as an abstract vector in Hilbert space, is essentially inevitable if we think that such an evolving quantum state provides a sufficient and complete description of physical reality.
Our approach is to take the quantum state as fundamental and search for an appropriate classical limit, rather than quantizing any particular classical model.

In this work we tackle this problem, building on previous work on deriving emergent spatial geometry from entanglement of a quantum state \cite{Cao:2016mst}.
There we investigated how a spatial metric (distance along curves) could be derived from a quantum state using the mutual information between different factors in Hilbert space.
Our interest here is dynamical rather than static: to model the universe as a quantum state evolving in Hilbert space, show how the geometry of spacetime can emerge from the entanglement features of such a state in an appropriate factorization, and derive Einstein's equation in the semiclassical limit, an approach we label Bulk Entanglement Gravity (BEG).
This requires us to consider a somewhat generic quantum system (or operator algebra), and examine which properties of a complex quantum system may be important in emerging spacetime geometry and gravity.

To concretely implement aspects of BEG, we will restrict ourselves in this paper to quantum states corresponding to emergent spacetimes in the weak-field regime, small perturbations of Minkowski space.
(Since boundary conditions play no role in our analysis, the results will apply equally well to spacetimes with a nonzero cosmological constant, as long as we consider regions much smaller than the background curvature scale.)
This represents a significant departure from the AdS/CFT version of the geometry-from-entanglement program. 
That approach may be thought of as ``maximally holographic,'' with all of the bulk data encoded directly on the conformal boundary, and in particular the entanglement from which geometry emerges is that of the CFT state.
Our regime is ``anti-holographic," considering a small region that is a weak perturbation of flat spacetime.
We are therefore interested in the entanglement of quantum states in Hilbert spaces that can be decomposed directly into factors corresponding to local regions of bulk spacetime (or equivalent ways of encoding entanglement data).

We will consider three different aspects of the BEG program. 
The first involves deriving emergent spatial geometry from entanglement data.
Using assumptions \ref{itm:1}, \ref{itm:2} and \ref{itm:3} below, one may start from an appropriate quantum state and obtain an emergent geometry and its best-fit dimensionality. 
Rather than deriving \emph{distances} from mutual information, it is more natural to derive \emph{areas} of surfaces using the entanglement entropy across them.
We will show in section \ref{sec:emergentspace} that for some class of geometries, the metric tensor can be obtained from these data using the tensor Radon transform \cite{sharafutdinov1994integral}. In the case of AdS/CFT, a procedure like this would correspond to directly recovering the bulk geometry (plus matter) from a state of the fundamental theory, such as that of a CFT, but without relying on the knowledge that it has a flat geometry or that it can be interpreted to reside on the asymptotic boundary of the emergent geometry. In general, we still refer to this emergent spatial geometry as the ``bulk'' geometry, even when no boundary theory is available. 

The second aspect is the emergence of gravitational dynamics.
In section \ref{sec:emergentgravity}, we show that the linearized Einstein's equation can be derived from a background-free approach using quantum entanglement when a set of assumptions outlined in the next section are satisfied. 
In particular, one can derive the Hamiltonian constraint \ref{subsec:MEECHamiltonianConstraint} from assumptions \ref{itm:1} through \ref{itm:5}. 
Our approach is closely related to the entanglement-equilibrium proposal of \cite{Jacobson:2015hqa}. It differs from \cite{Jacobson:2015hqa} in that the analogous entanglement condition is valid across global cuts, instead of small local spherical surfaces. In addition, it is valid for all matter fields and does not rely on CFT modular Hamiltonians, which require matter fields to have UV fixed points.
The main difference is that we derive our results directly from an abstract quantum state, rather than starting with quantum fields on an existing classical spacetime. 

The third aspect deals with the ``emergence map'': the map from abstract quantum states in Hilbert space to quantum fields on a semiclassical background geometry.
Part of this task can be thought of as determining which quantum degrees of freedom are responsible for emergent geometry as opposed to matter fields. This is important for the emergence of Einstein's equation with sources. 
To this end, we elaborate the relations with entanglement equilibrium and the Ryu-Takayanagi (RT) formula in section \ref{sec:generalizeRT}. In particular, we will discuss how one can distinguish the semiclassical geometry from quantum fields on that geometry solely from the features in a general background-free setting. Following Harlow \cite{Harlow:2016vwg}, we argue that this can be done purely from the entanglement structure of the state or from the properties of a quantum error-correction code (QECC). 
It seems that quantum error correction properties can naturally provide a separation between geometric and matter degrees of freedom. 

Our framework, partly inspired by \cite{Czech:2016tqr}, uses the Radon transform to tie together previous work on the thermodynamics of spacetime \cite{Jacobson:1995ab,Jacobson:2015hqa}, AdS/CFT approaches to emergent gravity \cite{Faulkner:2013ica}, and kinematic space \cite{Czech:2015qta}. 
The Radon transform can also be used to construct emergent geometries for quantum error correction codes or tensor networks in general. 

For the sake of concreteness, we will use the specific language of entanglement as computed from a quantum state in a Hilbert space. However, this work only relies on a configuration that defines subsystems and entanglement entropy data. Consequently, it also applies to more general formulations.

\section{The Road to Bulk Entanglement Gravity}

Our derivation of Einstein's equation from entanglement in the bulk of spacetime can be considered axiomatically: we can specify a list of explicit assumptions allowing us to start with an abstract quantum state and derive a semiclassical spacetime geometry with the appropriate dynamics.
Here we briefly list the assumptions, before discussing them in detail in subsequent sections.
Some of these assumptions will seem \emph{prima facie} reasonable, while others are more conjectural.
We will present arguments for their validity where available, and sketch a roadmap for the future work to complete this kind of program.

Our assumptions are as follows:
\begin{enumerate}[label=\textnormal{(A\arabic*)}]
\item \emph{Factorization.} Hilbert space $\mathcal{H}$ comes equipped with a preferred tensor product decomposition into individual factors,
 \begin{equation}
 \mathcal{H}=\bigotimes_i \mathcal{H}_i.
 \label{eqn:tensordecomp}
 \end{equation} 
 The individual factors $\mathcal{H}_i$ will correspond roughly to local points or small regions of space.
 The decomposition may be defined by the dynamics of the theory, as in \cite{Cotler:2017abq}.
 \label{itm:1}
\item \emph{Redundancy constraint.} We are given a state $|\Psi\rangle$ in this decomposition with a very specific behavior of the entanglement entropy: the entanglement entropies of individual factors (or groups thereof) are approximately ``redundancy constrained'' (RC). 
Given a collection $\textbf{B}$ of factors of $\mathcal{H}$ and its complement $\bar{\textbf{B}}$, a state is RC if its entropy can be written as a sum over the mutual informations of the individual factors,
  \begin{equation}
  S(\textbf{B}):=\frac 1 2 \sum_{i\in\textbf{B},j\in\bar{\textbf{B}}} I(i\co j).
 \label{eqn:RC}
 \end{equation}
 (Details of our notation are given in Section~\ref{sec:briefreview}.)
 Thus, RC states generalize the notion of area-law states.
 In an approximately RC state, the entanglement entropy of a subsystem $\mathbf{B}$ can be written as 
 \begin{equation}
 S(\mathbf{B}) = S_{\rm RC} +S_{\rm sub},
 \label{eqn:approxRC}
 \end{equation}
 where $S_{\rm RC}$ is the leading order contribution that satisfies the RC condition, and $S_{\rm sub}$ is a subleading correction. \label{itm:2}
\item \emph{Area from mutual information.} For states that define an emergent geometry, the mutual information $I$ between subsystems is proportional to the interface area $\area$ between corresponding subregions in that geometry, \label{itm:3}
\begin{equation}
 {\area}(\mathbf{B}, \bar{\mathbf{B}}) = \frac{1}{2\alpha} I(\mathbf{B} \co  \bar{\mathbf{B}}).
\end{equation}
 \item \emph{Entanglement equilibrium.} Entanglement perturbations of this configuration satisfy a modified entanglement equilibrium condition (MEEC), following Jacobson \cite{Jacobson:2015hqa}.
That is, under small perturbations, the total entropy perturbation $\delta S(R)$ of certain subsystems vanish, so that \label{itm:4}
\begin{equation}
  0 = \delta S_{\mathrm{RC}} +\delta S_\mathrm{sub}.
\end{equation}
 \item \emph{Emergent field theory.}
 The variation of the subleading correction can be generated by the entanglement variation of a state in some emergent effective field theory (EFT), \label{itm:5} 
 \begin{equation}
 \delta S_{\rm sub}=\delta S_{\rm EFT}.
 \end{equation}
Here by $S_{\rm EFT}$ we mean the vacuum-subtracted or Casini entropy, representing entanglement over and above the divergent contribution of the QFT vacuum \cite{Casini:2008cr,Bousso:2014sda,Bousso:2014uxa}.
 \item \emph{Dynamics.}
 There exists a consistent dynamical theory, e.g., a Hamiltonian or a quantum circuit, that generates a sequence of such configurations, each admitting an emergent spatial geometry. Furthermore, there is a  way to organize these emergent geometries to create a consistent Lorentzian spacetime geometry via time evolution. 
\label{itm:6}
 \item \emph{Lorentz invariance.}
The above assumptions hold for any constant-time slice of the emergent Minkowski space, and the overall theory is Lorentz-invariant in an appropriate limit. \label{itm:7}
\end{enumerate}

In the sections that follow we will show how to weave together these assumptions to derive geometry and Einstein's equation in the weak-field regime.

\section{Emergent Spatial Geometries and Radon Transforms}\label{sec:emergentspace}

In our earlier work on emergent space \cite{Cao:2016mst} we derived a spatial metric by using the quantum mutual information of two factors of Hilbert space to define a distance measure, based on our intuition from quantum field theory that the entanglement of low-energy states decreases monotonically with distance.
If entanglement is our main quantity of interest, however, it is more natural to directly derive areas from the entanglement across a boundary separating two regions, rather than to derive distances between any two small regions. 
We expect the entropy across such a boundary to be proportional to the geometric area, plus some subdominant correction.
In this section we explore the Radon transform as a natural tool for characterizing this data, and converting the entanglement of a quantum state into the metric tensor of a spatial slice.

\subsection{Space from Hilbert Space}\label{sec:briefreview}

Here we briefly review the emergence of spatial geometry from appropriate quantum states \cite{Cao:2016mst}.
Following assumption \ref{itm:1}, we are given a quantum state and a tensor product decomposition of the Hilbert space, $|\psi\rangle\in\mathcal{H}=\bigotimes_i\mathcal{H}_i$. 
The individual factors $\mathcal{H}_i$ correspond roughly to the degrees of freedom (geometric and field-theoretic) associated with a small local region of space.
In the back of our minds we are thinking of these factors as finite-dimensional vector spaces \cite{Bao:2017rnv}, though this doesn't play a crucial role in our analysis.
While this local picture of degrees of freedom runs against the spirit of holography, our interest here is in the weak-gravity regime, where it should be sufficient to think of gravity as a theory of local degrees of freedom.

\begin{figure}
 \includegraphics[width=0.65\textwidth]{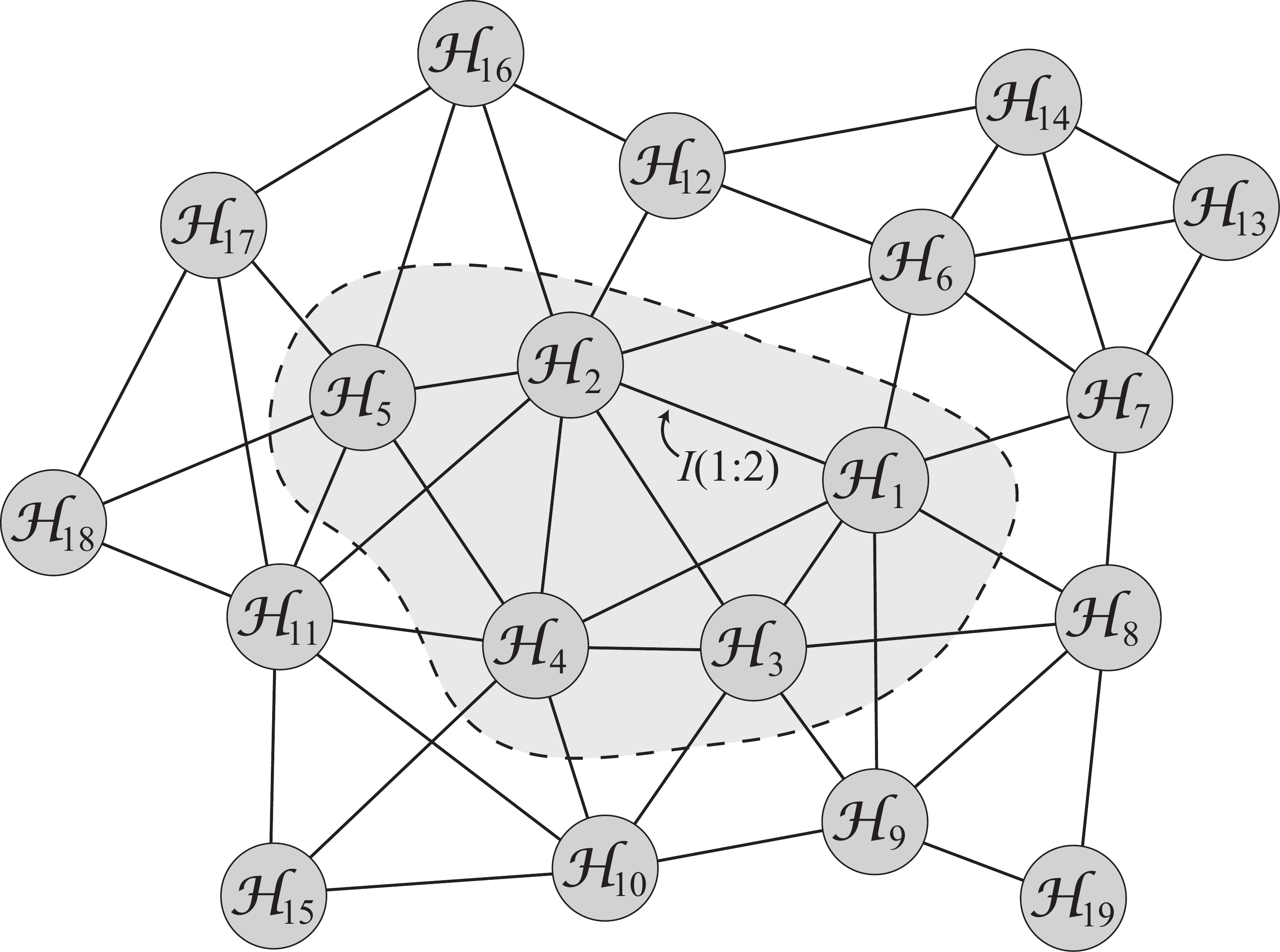}
\caption{This shows an example of the information graph in which vertices represent factors in a decomposition of Hilbert space, and edges are weighted by the mutual information between the factors. In redundancy-constrained states, the entropy of a group of factors (such as the shaded region ${\bold B}$ containing $\mathcal{H}_1\otimes\mathcal{H}_2\otimes\mathcal{H}_3\otimes\mathcal{H}_4\otimes\mathcal{H}_5$.) can be calculated by summing over the mutual information of the cut edges, as in (\ref{eqn:cutfunction}).
}
\label{fig:regions}
\end{figure}

One can generate an ``information graph'' $G=(V,E)$ based on this structure. The graph vertices in $V=\{i\}$ label each individual Hilbert space factors in $\{\mathcal{H}_i\}$, and the edges $E$ between any two vertices $i,j$ are weighted by the quantum mutual information between those factors,
\begin{equation}
  I(i\co j)=S(i)+S(j)-S(i\cup j).
\end{equation}
An example graph is shown in  Figure~\ref{fig:regions}.
For convenience, when we talk about quantities associated with the quantum state or the Hilbert space, we will use graph vertices and sets of vertices to denote tensor factors of the Hilbert space and products of the tensor factors, respectively. 
Note that we are not given the graph as a fundamental piece of information; it is derived from the quantum state in this particular factorization.
The factorization itself could be derived, for example, from the requirement that the Hamiltonian look approximately local \cite{Cotler:2017abq}.

We say that the state $|\psi\rangle$ (or more generally, the entanglement data) is ``redundancy constrained'' (RC) if the entanglement entropy of any subsystem $\textbf{B}\subset V$ can be computed by summing over the weights of all edges that connect vertices in $\textbf{B}$ with those in its complement $\bar{\textbf{B}}$, as in assumption \ref{itm:2}. More precisely, the entanglement entropy of a subsystem $\textbf{B}$ in a redundancy-constrained state is given by the cut function \cite{Bao:2015boa},
\begin{equation}
 S(\textbf{B})=S_{\rm RC}(\textbf{B}):=\frac 1 2 \sum_{i\in\textbf{B},j\in\bar{\textbf{B}}} I(i\co j).
 \label{eqn:cutfunction}
\end{equation}
Other than the familiar examples, such as area-law states  \cite{AreaLawEntRev} in certain condensed matter systems, a wider class of states such as Projected Entangled-Pair states \cite{Orus:2013kga}, holographic quantum error correction code \cite{Pastawski:2015qua}, and (bulk) random tensor network states \cite{Hayden:2016cfa} are also (approximately) redundancy constrained.
Note that a \emph{generic} state in Hilbert space will be very far from RC, and the information graph will be highly connected rather than taking the sparse form suggested by locality; the states we have in mind resemble low-energy states of approximately-local Hamiltonians.

In states that are only approximately RC, (\ref{eqn:cutfunction}) holds to leading order, and the exact entanglement entropy takes on a subleading correction
\begin{equation}
 S(\textbf{B})=S_{\rm RC}(\textbf{B})+ S_{\rm sub}(\textbf{B}),
\end{equation}
where $S_{\rm sub}\ll S_{\rm RC}$. 
There are two natural sources for such corrections.
One is long-range entanglement even in the vacuum, which we expect to be present but subdominant.
The other is entanglement between excited degrees of freedom over and above the vacuum.
An EPR pair, for example, can have an entanglement that is independent of the distance between the two particles; however in a quantum field theory such entanglement is a very small correction to the huge entanglement between the vacuum modes.
(For discussions of vacuum-subtracted entropy in quantum field theory, see \cite{Casini:2008cr,Bousso:2014sda,Bousso:2014uxa}.)

In order to obtain an emergent space, one has to make certain assumptions about the connection between entanglement and geometry. A natural identification can be motivated by area-law systems, where the entanglement entropy of a region scales as the interface area that separates the region and its complement. To leading order, this implies that the interface area is proportional to the mutual information between the region and its complement. Because RC states generalize area-law states, a natural definition is to define the ``interface area'' of an emergent geometry as the mutual information between a system and its complement, as in assumption \ref{itm:3}.
That is,
\begin{equation}
 \area(\textbf{B}, \bar{\textbf{B}}):=\frac{1}{2\alpha} I(\textbf{B}\co \bar{\textbf{B}}).
\end{equation}
See also \cite{Bianchi:2012ev,Myers:2013lva}.
At this level $\alpha$ serves as an undetermined constant of proportionality; we will later relate it to Newton's gravitational constant via $\alpha = 1/4G_N$.

Given the information graph, however, it is more convenient to work with the length measure between factors, instead of their mutual interface areas.
To derive the approximate geometry that may be encoded by the graph, \cite{Cao:2016mst} defines an \textit{ad hoc} distance function between vertices from the edge weights. This, together with the set $V$, generate a metric space, which is isometrically embedded in to a manifold. We define the embedding manifold as the emergent geometry of the graph. The technique of classical multi-dimensional scaling (MDS) \cite{cmds,walter,eladkimmel,gmds,mdsbook} can be used for this purpose to determine the best-fit dimensionality of the geometry, as well as the embedding coordinates for the elements of $V$.

\subsection{Metric tensor from the inverse tensor Radon transform}
\label{sec:metricfromradon}

While straightforward in an approximate recovery of geometry, the correct ``transform'' function from an area quantity to distances can, in principle, be non-local. A simple local function from area to distance is expected to yield a distortion unless we operate in the case with a high degree of symmetry. To address this deficiency, we seek an improved method to directly transform (dualize) the area quantity, which is proportional to mutual information, to a measure of distances via a global transformation. In this section, we reconstruct the spatial metric tensor directly from entanglement data.

In cases of interest where the emergent geometry is not highly symmetric, we can imagine a two-part procedure, in which we use MDS to emerge a symmetric ``background'' geometry, and then recover the metric perturbation from an inverse tensor Radon transform. 
(In fact the background geometries of interest to us will generally be flat Euclidean spaces.)
The recovery procedure is valid as long as the tensor Radon transform has a unique inverse. Such is the case for simple manifolds in 2 dimensions  \cite{BRP2dproof}, which have been extensively studied in the context of the boundary-rigidity problem. In higher dimensions, similar inversions are also possible in Riemannian geometries that are close to flat space or hyperbolic space \cite{uhlmann}.

\begin{figure}
 \includegraphics[width=0.75\textwidth]{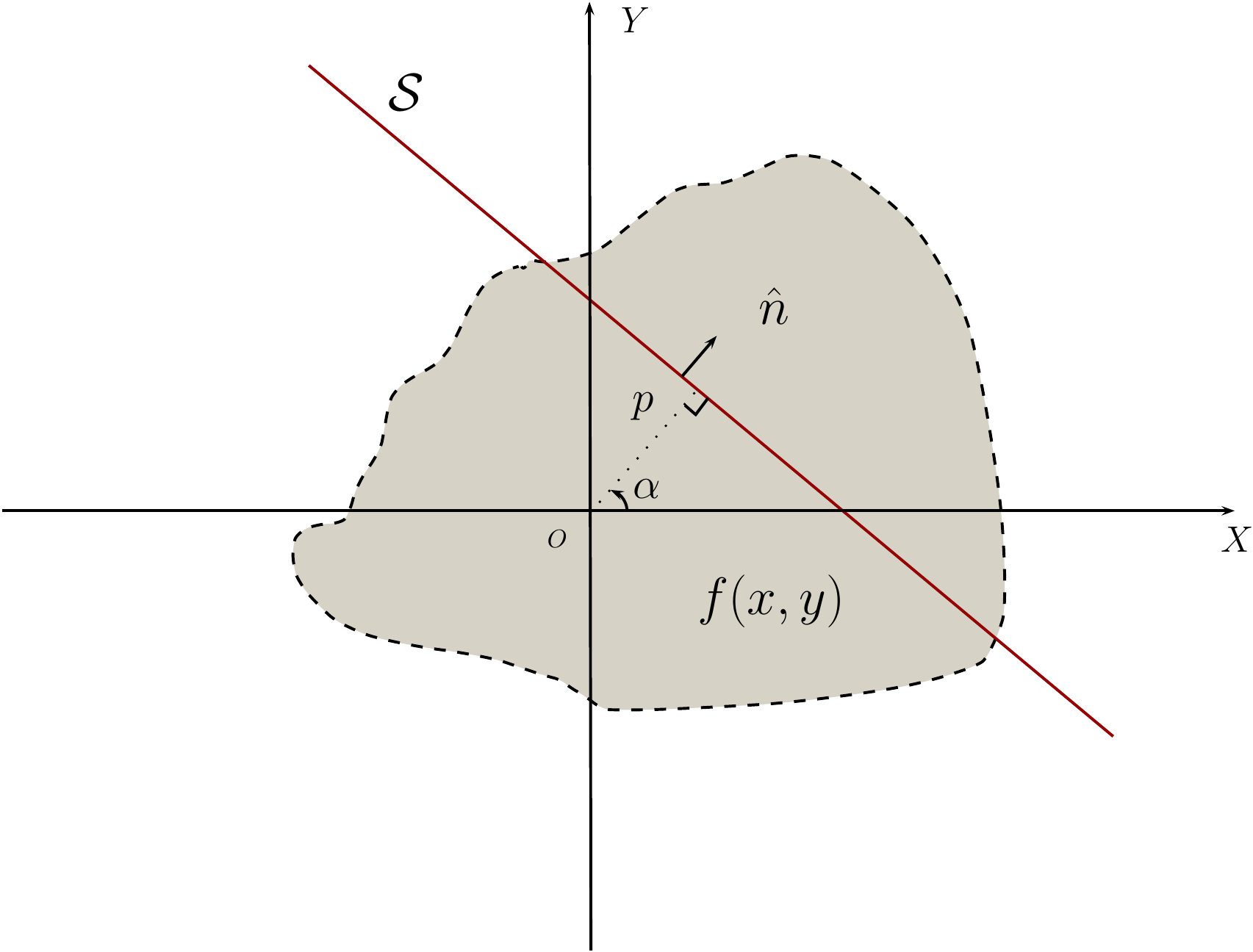}
\caption{The Radon transform of a scalar function $f(x,y)$ defined on some compact domain (shaded area) is done by integrating the function over the submanifold $\mathcal{S}$, which is a line in 2-dimensional flat space. The transformed data $\R[f](p,\alpha)$ corresponds to the value at a point in the space of $\mathcal{S}$ (the space of lines in this case). $p$ is the perpendicular distance between the plane and the origin and $\alpha$ parametrizes the direction of the unit normal $\hat{n}$.
}
\label{fig:xraytransform}
\end{figure}

Intuitively, the Radon transform maps a function on a space to a function on a set of surfaces embedded in that space, by integrating the function over the surface  \cite{helgason1999Radon,sharafutdinov1994integral,Czech:2016tqr}.
More formally, consider an $n$-dimensional Riemannian manifold $\mathcal{M}$ and a totally-geodesic codimension-1 submanifold $\mathcal{S}$, where ``totally geodesic'' means that the geodesics of the submanifold with respect to its induced metric are also geodesics of the original manifold.\footnote{Analogous transforms along $n-k$ dimensional submanifolds can also be defined. Here we only discuss the case when $k=1$.} 
Most geometries will not admit any totally-geodesic submanifolds, but they are plentiful in the highly-symmetric backgrounds of interest to us here, \emph{e.g.} hyperplanes in Euclidean spaces.
The Radon transform of a function $f$ on $\mathcal{M}$ at $\mathcal{S}$ is defined as the integral 
\begin{equation}
\R[f](\mathcal{S}) = \int_{\mathcal{S}} f\, d\sigma
\end{equation}
over $\mathcal{S}$ with area element $d\sigma$. 
Clearly, a well-defined transform requires the function to be regularized in some way, \textit{e.g.,} by setting $f=0$ outside some domain. 
If the geometry on $\mathcal{M}$ is Euclidean (in the sense of flat), an appropriate set of surfaces $\mathcal{S}$ is given by planes specified by a distance and angle from the origin, as shown in Figure~\ref{fig:xraytransform}.

We can also perform the Radon transform of a tensor field.
Such tensor Radon transforms were used in  \cite{Czech:2016tqr} to derive the linearized Einstein's equation in the context AdS/CFT. 
We will employ analogous techniques, but directly in the bulk, without reference to kinematic space or holography. 

Let $g_{ij}$ be the metric tensor on $\mathcal{M}$ and let $w_{ij}$ be the induced metric of the submanifold $\mathcal{S}$. 
(Our notation follows \cite{Czech:2016tqr}, and differs from the more common notation in the mathematical literature.)
The longitudinal tensor Radon transform of $s_{ij}$ on $\mathcal{M}$ is defined as
\begin{equation}
\R_{\parallel} [s_{ij}] = \int_{\mathcal{S}} w^{ij} s_{ij}\, d\sigma.
\label{eq:Rparallel}
\end{equation}
(Henceforth we will drop the explicit appearance of the submanifold $\mathcal{S}$ on the right-hand side.)
Similarly, the transverse tensor Radon transform is
\begin{equation}
\R_{\bot}[s_{ij}] = \int_{\mathcal{S}} (g^{ij}-w^{ij}) s_{ij}\, d\sigma.
\end{equation}
Indices are raised and lowered with the spatial metric. Since at this point we are only considering spatial geometry and related tensors, we do not discuss metrics with Lorentzian signatures. 
A process to invert the above transform and obtain the tensor $s_{ij}$ is referred to as inverse tensor Radon transform. 

The inversion problem in $n=2$ has been mostly studied in the context of the boundary-rigidity problem  \cite{BRPrev1,Stefanov2008}, which examines if the bulk geometry of a manifold can be recovered knowing only the pair-wise geodesic distances between all its boundary points. A manifold for which this is possible is called boundary-rigid. 
This problem has been shown \cite{sharafutdinov1994integral} to be equivalent to the tensor geodesic X-ray transform problem \cite{john1938,sharafutdinov1994integral, Stefanov2008}, which coincides with the tensor Radon transform in 2 dimensions. An earlier classification of boundary-rigid manifolds is now known as the Michel's conjecture \cite{michel}, where so-called simple manifolds are boundary-rigid. (A Riemannian manifold $\mathcal{M}$ is simple if, given any two points, there exists a unique geodesic joining the points, and if the second fundamental form is positive definite at every point on $\partial\mathcal{M}$.) A proof has been given in $n=2$ \cite{BRP2dproof}, although some other higher-dimensional results are also known \cite{BRPflat}. See  \cite{Stefanov2008,BRPrev2,Croke2004} and references therein. Results related to applying the inverse were explored both analytically \cite{Sharafutdinov2007,Stefanov2014} and numerically \cite{Monard}. 
The inverse problem of the higher-dimensional tensor Radon transform has remained largely unexplored until recently, where a proof on invertibility was produced \cite{uhlmann}, but an explicit inversion formula and numerical results are still unknown, to the best of our knowledge. 
Henceforth we will simply assume that the appropriate tensor-transform inversion can be performed.

\subsection{Spatial metric from entanglement}

We now describe how to use the Radon transform to obtain an emergent spatial geometry from the quantum state.
Suppose we begin with a quantum state $|\psi\rangle$ from which, following \cite{Cao:2016mst}, we may use MDS to find a best-fit maximally-symmetric geometry $g_{ij}$ on a spatial manifold $\{\mathcal{M}\}$, which we will refer to as the background.
We now would like to consider a perturbed state,
\begin{equation}
  |\Psi\rangle = |\psi\rangle + \delta |\psi\rangle.
\end{equation}
The perturbed entanglement entropy $\delta S$ associated with each subsystem also changes the associated emergent geometry. 
Using these perturbed data $\delta S$ for different subsystems, we will show that the inverse tensor Radon transform allows us to recover a perturbed metric $\delta h_{ij}$ on the background.\footnote{A similar procedure may be used when MDS itself gives an imperfect embedding of the background geometry.}  

We are thus considering a situation in which we attempt to recover $\delta h_{ij}$ when we are given a background metric $g_{ij}$ and an entanglement perturbation $\delta S$ that can be computed from the state itself. 
For discrete finite-dimensional quantum systems, such as some condensed matter models \cite{ising1,ising2, Kitaev97} or the ones we considered in  \cite{Cao:2016mst}, we will also assume a continuum limit or a smoothing process over the data $\delta S$ such that the usual Radon transform is well-defined and can be performed. Alternatively, a discrete version may also be applied  \cite{1165108}. 

We illustrate the reconstruction of $\delta h_{ij}$ with an example in flat space, although the procedure can be easily generalized to other backgrounds, as long as an inverse transform exists. Consider the case where one determined the  exact geometry encoded in the state $|\psi\rangle$ to be an $n$-dimensional flat space with metric $g_{ij}=\delta_{ij}$. For any codimension-1 hyperplane $\mathcal{C}(p,\hat{n})$ that separates the space into two adjacent regions ${\region}, \bar{\region}$, one can compute the interface area $\area(\mathcal{C})$ using the flat metric of the  embedding space. This follows from assumption \ref{itm:3}, that area is proportional to mutual information,\footnote{With MDS, we are considering only a geometry that is finite in extent. In the case where the space is infinite, we restrict ourselves to a particular finite region for analysis. }
\begin{equation}
\area(\mathcal{C})=\frac{1}{2\alpha} I({\region}_{\mathcal{C}}\co \bar{{\region}}_{\mathcal{C}}).
\label{eqn:aptmi}
\end{equation} 

Given the RC assumption \ref{itm:2}, $I({\region}_{\mathcal{C}}\co\bar{{\region}}_{\mathcal{C}})$ is determined by the sum of mutual informations along all edges that are cut by $\mathcal{C}$ \cite{Bao:2015boa}. 
Adding a perturbation $\delta|\psi\rangle$ in general also perturbs the mutual information across different bipartitions. For the same bipartition along the cut, the perturbed area is now given by perturbed mutual information $\area' = (1/2\alpha)I'({\region}_{\mathcal{C}}\co \bar{{\region}}_{\mathcal{C}})$, and one can define the area perturbation 
\begin{equation}
\delta \area =\area'-\area= \frac{1}{2\alpha} \delta I =  \frac{1}{2\alpha} (I'-I)({\region}_{\mathcal{C}}\co\bar{{\region}}_{\mathcal{C}}),
\end{equation}
so that
\begin{equation}
\delta I(\mathcal{C})=2\alpha\delta \area (\mathcal{C}).
\end{equation}

Let $\tilde{w}_{ij}$ be the induced metric of $\mathcal{C}$ in the perturbed geometry, where $w_{ij}$ is the induced metric in the background,
\begin{equation}
  \tilde{w}_{ij} = w_{ij} + \delta w_{ij}.
\end{equation}
The area is then
\begin{equation}
  \area'(\mathcal{C}) = \int_\mathcal{C} \sqrt{\det{\tilde{w}_{ij}}}\, d\sigma,
\end{equation}
and the area perturbation is 
\begin{equation}
  \delta \area = \int_\mathcal{C} \left(\sqrt{\det{\tilde{w}_{ij}}}-\sqrt{\det{{w}_{ij}}}\right)\, d\sigma.
\end{equation}
Using $\det(\mathbb{I}+\epsilon M)=1+\epsilon\Tr[M]+O(\epsilon^2)$ for any symmetric matrix $M$, this becomes
\begin{equation}
  \delta \area = \frac 1 2 \int_\mathcal{C} \Tr (\delta w) \, d\sigma,
\end{equation}
where $\Tr (\delta w) = w^{ij}\delta w_{ij}$.
Comparing to (\ref{eq:Rparallel}), we see that the area perturbation is directly related to the longitudinal Radon transform of the induced metric perturbation,
\begin{equation}
  \delta \area = \frac 1 2\R_\parallel [\delta w_{ij}].
\end{equation}
It is straightforward to show, for example by choosing an appropriate coordinate system, that $\R_{\parallel}[\delta w_{ij}] =\R_{\parallel}[\delta h_{ij}] $.
We therefore see that the mutual information across the bipartition is proportional to the Radon transform of the metric perturbation,
\begin{equation}
\delta I(\mathcal{C})= \alpha \R_{\parallel}[\delta h_{ij}] .
\end{equation}
Given the entanglement data $\delta I(\mathcal{C}) =2\alpha \delta \area(\mathcal{C})$ over all such cuts $\mathcal{C}(p,\hat{n})$, we can perform the inverse tensor Radon transform of $ \R_{\parallel}[\delta h_{ij}]$, thus completing the metric reconstruction procedure, so that the full spatial metric $g_{ij}+\delta h_{ij}$ is obtained from entanglement data of a quantum state in a background-free approach. 

We therefore need assurance that the tensor Radon transform of interest is indeed invertible.
At first sight, this requirement of invertibility to recover a tensor from a scalar function seem unlikely, simply from counting degrees of freedom; there are several components of the metric, and only one value each of $\R_\bot$ and $\R_\parallel$. 
However, this is indeed uniquely invertible for a certain set of manifolds, up to natural obstructions that are not simply fixed by the data. In this case, the degrees of freedom that are undetermined by entanglement are manifested as gauge transformations of the $\delta h_{ij}$ field by an arbitrary vector field $\xi$,
\begin{equation}
\delta h_{ij} \rightarrow \delta h_{ij} +\partial_i \xi_j +\partial_j\xi_i,
\end{equation}
for the simple reason that the longitudinal tensor Radon transform vanishes for tensors of the form $\partial_{(j}\xi_{i)}$ .

From the existing mathematical literature, we conclude that for $n=2$, an inverse transform can be explicitly implemented to obtain $\delta h_{ij}$ from an entanglement perturbation as long as the background manifold $\mathcal{M}$ is boundary-rigid. Current knowledge classifies simple manifolds and certain quotients  \cite{finitequotient} as boundary-rigid, although other particulate examples such as tori have also been given  \cite{Croke2004,torus2,croke1991}. For $n>2$, an inverse, if it exists, can also be uniquely obtained near flat or hyperbolic geometries. Recently, \cite{Tracey2017} also proposes a recovery of the metric for certain types of Riemannian manifolds at $n=3$. However, an explicit reconstruction algorithm for general dimensions is still contingent on further progress in the mathematical community.\footnote{The boundary-rigidity problem is intimately related to that of reconstructing bulk geometry from boundary data in the context of AdS/CFT. The existence of a manifold that is not boundary-rigid may be indicative of the limitations of the Ryu-Takayanagi formula in $\rm AdS_3/CFT_2$ for constructing bulk geometries. Similar conclusions also apply to certain reconstruction schemes in higher dimensions, which use correlation functions to estimate geodesic lengths  \cite{Porrati:2003na}. For instance, consider the back-reacted geometry of a single massive particle in AdS. The spatial geometry of a time slice is not boundary-rigid.}

\section{Emergent Gravity from Quantum Entanglement}\label{sec:emergentgravity}

We now turn to the gravitational dynamics of our emergent geometries, and explain how assumptions \ref{itm:1} through \ref{itm:7} allow us to derive the linearized Einstein equation in the weak-field limit.
In the first subsection we look at classical spacetime, using the Radon transform to write the terms appearing in Einstein's equation in a convenient form.
In the following subsection we use the results to derive a modified entanglement equilibrium condition, and in the final
subsection we derive Einstein's equation for an emergent Lorentzian spacetime geometry from the quantum state.

\subsection{The Hamiltonian Constraint and its Radon transform}\label{subsec:MEECHamiltonianConstraint}

Let us momentarily set aside spacetime emerging from quantum mechanics and instead consider the conventional classical Einstein's equation linearized around a Minkowski background. 

Given a spacetime with a parameterized set of time slices $\mathcal{M}_t$ with timelike unit normal vectors $t^\mu$, the classical Hamiltonian constraint of general relativity corresponds to the condition
\begin{equation}
G_{\mu\nu}t^{\mu}t^{\nu}= 8\pi G_N T_{\mu\nu}t^{\mu}t^{\nu}.
\end{equation}
In the following derivation, we work in the linearized regime where we consider metric perturbations on a Minkowski background: $g_{\mu\nu} = \eta_{\mu\nu}+\delta h_{\mu\nu}$. As such, we can consider the constant-time slices of Minkowski space.
The (background) extrinsic curvature vanishes on each of these time slices. 
At the linearized level we can then relate the Einstein tensor to the \emph{spatial} curvature scalar via  $\mathcal{R} = 2G_{tt}$ \cite{Gourgoulhon:2007ue}.
Therefore, for linearized equations, the Hamiltonian constraint reads
\begin{equation}
 \delta \mathcal{R} = 16\pi G_N \delta T_{tt}.
\label{eqn:HC2}
\end{equation}

Because we are specializing to flat backgrounds, for each such constant time slice, let the background spatial metric be $\delta_{ij}$ and the perturbation be $\delta h_{ij}$, which is a tensor-valued function on flat space. 
To linear order we can expand 
\begin{equation}
 \delta \mathcal{R}=\delta^{ij}\frac 1 2 (\partial_i\partial_r \delta h_{jr}+\partial_j\partial_r \delta h_{ir}-\partial_i\partial_j\delta h-\nabla^2 \delta h_{ij}),
 \label{eqn:Rexpand}
\end{equation}
where $\nabla^2$ is the Laplacian in $n$-dimensional flat space, with $n$ being the dimension of the constant-time slice. 
Now consider taking the Radon transform of both sides of equation (\ref{eqn:Rexpand}) along $(n-1)$-dimensional hyperplanes. We can derive\footnote{The derivation is simpler by writing the metric in the Gaussian normal coordinate.}
\begin{equation}
 \R[\delta \mathcal{R}]=- \R_{\parallel}[\nabla^2 \delta h_{ij}].
\end{equation}
Radon transforms obey an intertwinement relation \cite{helgason1999Radon},
\begin{equation}
  \R[\nabla^2 f] = \frac{\partial^2}{\partial p^2} \R[f],
\end{equation}
where $p$ is the distance from the origin to the hyperplane.
From this and the fact that $\R_{\parallel}[\delta h_{ij}]=2\delta \area$, we find a relation between the spatial curvature and the area perturbation,
\begin{equation}
 \R[\delta \mathcal{R}] =- 2\frac{\partial^2}{\partial p^2}\delta \area.
\end{equation}
Comparing to (\ref{eqn:HC2}), we end up with
\begin{equation}
-\frac{\partial^2}{\partial p^2}\delta \area = 8\pi G_N \R[\delta T_{tt}].
\label{eqn:Radonricci}
\end{equation}

To solve (\ref{eqn:Radonricci}) for $\delta \area$, we convolve the source with the Green function,\footnote{The full Green function has additional terms $c_1p+c_2$. However, since the boundary condition is unknown, unlike the case in AdS/CFT, we fix the coefficients by requiring the solution $\delta \area$ matches to the Rindler Hamiltonian of the bulk matter fields, by choosing $c_1=c_2=0$. }
\begin{equation}
G(p,q) = (q-p)\theta(q-p).
\end{equation}
This yields
\begin{align}
\delta \area &=- \int  (q-p)\theta(q-p) \R[8\pi G \delta T_{tt}] \, dq\\
&=-8\pi G_N\int_{\mathcal{C}}\int_{q>p} (q-p)\delta T_{tt}\, dq\, d^{n-1}\sigma.
\label{eqn:areasoln}
\end{align}
Recall that each surface  $\mathcal{C}$ is specified by a distance parameter $p$ from the origin and its unit normal $\hat{n}$. Then the integral is over the half space up to a surface with unit normal $\hat{n}$ and distance $p$ from the origin. 

\subsection{Emergent entanglement equilibrium}

We can now connect these classical GR concepts to entanglement data of an underlying quantum state in an abstract Hilbert space.
Although the physical meaning of equation (\ref{eqn:areasoln}) is unambiguous in the classical theory, these quantities should ultimately be derived from the quantum state if space is emergent from entanglement. On the left-hand side, as we know from previous constructions \cite{Cao:2016mst}, the area perturbation $\delta A$ can be identified with the sum of mutual information $\delta I/2\alpha$ along the cut $\mathcal{C}$ in an RC state. In the case where the overall state is pure, 
\begin{equation}
\delta I({\region}_{\mathcal{C}}\co \bar{{\region}}_{\mathcal{C}}) = 2\delta S({\region}_{\mathcal{C}}).
\end{equation}

On the right-hand side of (\ref{eqn:areasoln}), in a semiclassical theory we interpret the classical quantity $\delta T_{tt}$ as the expectation value of a quantum operator $\widehat{T}_{tt}$ in some particular state of a quantum field theory on curved spacetime.
We then recognize that the integral in (\ref{eqn:areasoln}) is related to the modular Hamiltonian of the right Rindler wedge for a quantum field theory, translated spatially by $p$. More explicitly, take $p=0$ and identify the normal of $\mathcal{C}$ with the direction in which the Rindler observer accelerates, which we take to be the $x$-axis.
We then have the QFT expression for the Rindler modular Hamiltonian,
\begin{equation}
\widehat{H}_{\rm mod}=2\pi\int\int_{x>0} x \widehat{T}_{tt} \, d^n x.
\end{equation}
We therefore consider the right-hand side of (\ref{eqn:areasoln}) to represent the expectation value of the modular Hamiltonian of some effective field theory on a flat background, evaluated with respect to some linearized perturbation of a quantum state $\delta \rho_{\rm EFT}$, such that 
\begin{equation}
\delta T_{tt}= \Tr[\delta\rho_{\rm EFT}\widehat{T}_{tt}]. 
\end{equation}
In the linearized regime, it must also be proportional to the entanglement entropy perturbation $\delta S_{\rm EFT}(\mathcal{C})$ of the same half-space demarcated by $\mathcal{C}$, via the entanglement first law \cite{Blanco:2013joa},
\begin{equation}
\delta\langle \widehat H_{\rm mod}\rangle=\delta S_{\rm EFT}(\mathcal{C}).
\end{equation}
Substituting these new variables and using (\ref{eqn:aptmi}), (\ref{eqn:areasoln}) becomes
\begin{equation}
\frac 1 2\delta I({\region}_{\mathcal{C}}\co\bar{{\region}}_{\mathcal{C}}) +4G\alpha\delta S_{\rm EFT}(\mathcal{C})=0 ~~\rm (pure~state),
\end{equation}
or using the RC relation (\ref{eqn:RC}), 
\begin{equation}
 \delta S({\region}_{\mathcal{C}})+4G_N\alpha\delta S_{\rm EFT}(\mathcal{C})=0.
\label{eqn:EEEC}
\end{equation}

Let us try to understand this relation in the context of deriving geometry from a quantum state. By construction, $\delta S({\region}_{\mathcal{C}})$, which is proportional to the area perturbation, is the contribution that we have consistently identified with the RC part of the entanglement. 
The more difficult question is how $\delta S_{\rm EFT}$ should be identified.
Recall that from assumption \ref{itm:2}, in ground states of systems satisfying an area law \cite{AreaLawEntRev} or other approximately-redundancy-constrained states, one can write the total entanglement entropy associated with a subsystem ${\region}$ as 
\begin{equation}
S_{\rm total}({\region}) = S_{\rm RC}({\region}) + S_{\rm sub}({\region}),
\label{eqn:entropysum}
\end{equation}
where the RC contribution $S_{\rm RC}$, or the area-scaling contribution when there is a well-defined geometry, dominates over the subleading correction $S_{\rm sub}$. 
 Motivated by the RT formula with subleading corrections, we claim in assumption \ref{itm:5} that $S_{\rm EFT}$ can be interpreted as originating from the subleading corrections $S_{\rm sub}$ to the RC entropy contribution. We will further discuss this claim and its similarities with a generalized form of the RT formula in section \ref{sec:generalizeRT}.

Jacobson \cite{Jacobson:2015hqa} derived Einstein's equation in a semiclassical bulk spacetime from the concept of entanglement equilibrium. This is the assumption that the total entanglement entropy of a small ball in some maximally-symmetric background spacetime is extremal, i.e., $\delta S_{\rm total} =0$ when a small perturbation is added. To complete the derivation, one has to separate the entanglement into UV and IR contributions, such that $\delta S_{\rm total}= \delta S_{\rm UV}+ \delta S_{\rm IR}$. In  \cite{Jacobson:2015hqa}, $\delta S_{\rm UV}\sim \delta A$ is assumed to be the area variation in some background geometry and $\delta S_{\rm IR}$ is identified with the entanglement entropy of a field theory regulated in some way.

Our equation (\ref{eqn:EEEC}) relating geometric entropy to the entropy perturbation of an emergent EFT can be thought of as a version of the modified entanglement equilibrium condition (MEEC) from assumption \ref{itm:4}. 
The geometric term corresponds to the UV contribution, while the EFT (matter) term corresponds to the infrared,
\begin{align}
\delta S({\region}_{\mathcal{C}})&\leftrightarrow \delta S_{\mathrm{UV}} \leftrightarrow \delta S_{\mathrm{RC}}, \\
 4G_N\alpha\delta S_{\rm EFT}(\mathcal{C})& \leftrightarrow\delta S_{\mathrm{IR}} \leftrightarrow \delta S_{\rm sub},
\end{align}
and the condition (\ref{eqn:EEEC}) states that these sum to zero near the background. A crucial difference, however, is instead of entanglement across some small ball centered at a point, the condition now has to hold across all cuts $\mathcal{C}$ made by $(n-1)$-dimensional totally-geodesic submanifolds in the background space. This also differs from  \cite{Jacobson:2015hqa} in that one no longer has to rely on CFT modular Hamiltonians by assuming the special property of the matter field theory having an UV fixed point. The result holds for a generic QFT with the corresponding Rindler Hamiltonian. 

For now, we will proceed with the identification $\delta S_{\rm sub} = \delta S_{\rm EFT}$.  
This fixes the constant $\alpha = 1/4G_N$, the value required for the consistency of Einstein's equation and the holographic bound. 
Consequently, MEEC translates into a more general relation, whereby UV and IR portions of the entropy are identified not based on assumptions in semiclassical physics, but rather on the properties of quantum entanglement,
\begin{align}
0&=\delta S_{\rm total}\cr &=\delta S_{\rm RC}+ \delta S_{\rm sub} \cr & =\alpha\delta A +\delta S_{\rm EFT}.
\label{eqn:BEGMEEC}
\end{align}
With this identification, the above relation is a necessary condition for a state to have emergent properties consistent with general relativity at low energies. 

Therefore, by making the identification that the subleading entropy to RC with matter field entropy, similar to the vacuum-subtracted (Casini) entropy  \cite{Casini:2008cr,Bousso:2014sda,Bousso:2014uxa}, we establish an equivalence between the modified entanglement equilibrium condition (\ref{eqn:BEGMEEC}) and the Radon transform of the linearized Hamiltonian constraint (\ref{eqn:Radonricci}). This argument can also be used to generalize the result of  \cite{Faulkner:2013ica} in AdS/CFT to other (non-flat) geometries, as long as the function is invertible under Radon transform in the background Riemannian manifold.

\subsection{Linearized Einstein Equation from entanglement}

We can now put the picture together to derive dynamics for the emergent spacetime geometry, in a way similar to \cite{Jacobson:2015hqa}.
For the sake of convenience, let's assume that from our previous results one has already emerged a flat background geometry from MDS or tensor Radon transform techniques. Similar to the AdS/CFT case considered in \cite{Faulkner:2013ica}, we now wish to determine if the geometric deformation from entanglement perturbations responds in a way consistent with Einstein gravity. 
A similar conclusion can also be generalized to hyperbolic spaces using MDS with a best fit curvature parameter following the procedure of \cite{Czech:2016tqr}, but we will not consider that case here. 

Consider the quantum system from which flat space is emergent. For concreteness, the total system could be described by a quantum state $|\psi\rangle\in\mathcal{H} = \bigotimes_i \mathcal{H}_i$, as in assumption \ref{itm:1}. A subsystem is thus described by the reduced density operator associated with some Hilbert subspace. Any cut $\mathcal{C}$ that corresponds to a codimension-1 hyperplane in the emergent geometry will bipartition the system into two adjacent non-overlapping regions. One can compute the entanglement entropy for either region $\Sigma$, which reads $S(\Sigma) = S_{\rm RC}(\Sigma) +S_{\rm sub}(\Sigma)$, as in \ref{itm:2}. Now we add a perturbation to obtain $|\Psi\rangle = |\psi\rangle + \delta |\psi\rangle$. The perturbation will modify the entanglement, which in turn changes the emergent geometry, following the area perturbation defined by \ref{itm:3}. 
 
The MEEC assumption \ref{itm:4} relates perturbations in the RC and subdominant contributions to the entropy across $\mathcal{C}$,
\begin{equation}
 0=\delta S_{\rm RC}+\delta S_{\rm sub}.
\end{equation}
Using assumption \ref{itm:5} to relate the subdominant term to the vacuum-subtracted entropy of an effective field theory, the MEEC is equivalent to the (scalar) Radon transform of the classical Hamiltonian constraint linearized against a flat background, 
\begin{equation}
\R[\delta \mathcal{R}]= 16\pi G_N \R[\delta T_{tt}],
\label{eqn:RadonHamiltonianConstraint}
\end{equation}
as argued in the previous subsection.
Here $\delta \mathcal{R}$ is the spatial curvature perturbation and $\delta T_{tt}$ is the linear perturbation of the stress-energy associated with an effective field theory living on the background. 
Because this relation holds for all such cuts in the flat background space, equation (\ref{eqn:RadonHamiltonianConstraint}) uniquely determines the linearized Hamiltonian constraint,
\begin{equation}
\delta\mathcal{R}= 16\pi G_N\delta T_{tt},
\label{eqn:HamiltonianConstraint}
\end{equation}
provided the inverse is well-defined. 

Following assumption \ref{itm:6} about dynamics, we consider a sequence of states $|\Psi (t)\rangle$ labeled by a single parameter $t$, which together describe a Lorentzian spacetime.
The corresponding emergent spatial geometries can be thought of as embedded spacelike slices in a spacetime with coordinates in synchronous gauge.
In terms of a unit timelike vector field $t^\mu$ normal to these slices, the Hamiltonian constraint (\ref{eqn:HamiltonianConstraint}) can be written as 
\begin{equation}
\delta G_{\mu\nu}t^{\mu}t^{\nu}= 8\pi G_N \delta T_{\mu\nu}t^{\mu}t^{\nu}.
\end{equation}
Under the Lorentz-invariance assumption \ref{itm:7}, this must be valid for arbitrary normal $t^{\mu}$. 
We therefore have the full linearized Einstein's equation,
\begin{equation}
\delta G_{\mu\nu}= 8\pi G_N \delta T_{\mu\nu}.
\end{equation}
While the number of conjectural assumptions needed to reach the result is admittedly considerable, we find the path we've outlined to be a promising route to deriving bulk gravitational dynamics directly from the evolution of an abstract wave function in Hilbert space.

\section{Entanglement RT Formula and Quantum Error Correction}\label{sec:generalizeRT}

It would be useful to have a more systematic approach to decomposing an abstract quantum state into geometric (UV) and matter (IR) degrees of freedom.
In the previous section we proposed one such procedure, identifying $S_{\rm RC}= \alpha \area$ and $S_{\mathrm{sub}}=S_{\rm EFT}$ when the state is approximately RC and admits an emergent geometry. This identification also proposes a background-free way of understanding these ``UV'' and ``IR'' entropies purely from the characteristics of entanglement, which can be done for arbitrary quantum states. 

In this section, we will connect these observations with the Ryu-Takayanagi (RT) formula from AdS/CFT. We will also argue that more general emergent geometries can be assigned to quantum error correction codes, where the code subspace naturally separates the geometric and matter contributions to entanglement entropy. Our considerations here are tentative (even by the standards of the rest of the paper), and would require more elaboration to make precise. 

Let's first recall the RT entropy relation in the case of AdS/CFT with a subleading $N^0$ correction \cite{Faulkner:2013ana},
\begin{equation}
S_{\mathrm{CFT}}(A) =\frac{\mathcal{A}_{\mathrm{ext}}(A)}{4G_N} +S_{\rm bulk}.
\label{eqn:RTwCorr}
\end{equation}
Here, $S_{\mathrm{CFT}}(A)$ is the entanglement entropy of a subregion $A$ in the boundary field theory,  $\mathcal{A}_{\mathrm{ext}}(A)$ denotes the area of a bulk extremal surface homologous to $A$, and $S_{\rm bulk}$ is a correction representing contributions from bulk matter fields. 

The RT formula relates boundary quantities to bulk quantities in a holographic setting.
But there is an obvious analogy to the BEG relation
\begin{align}
S_{\rm total} &= S_{\rm RC}+S_{\rm sub}\\ &= \frac{\area(\mathcal{C})}{4G} +S_{\rm EFT}(\mathcal{C})
\label{eqn:generalRT}
\end{align}
for a cut $\mathcal{C}$ by a totally geodesic codimension-1 submanifold of an emergent spacetime. 
In this case, the MEEC is an infinitesimal version of (\ref{eqn:generalRT}), and can be interpreted as a perturbative version of an RT-like relation for which $\delta S_{\mathrm{CFT}}=0$. It may be possible to understand this relation from a more general perspective, in which AdS/CFT is a special case manifested as a duality. 

If we interpret the bulk AdS as an emergent entity from the boundary conformal field theory, we can think of the CFT to be the fundamental theory, from which we reconstruct a theory in the IR that describes bulk gravity. In this emergent limit, different parts of the entanglement entropy of the supposed fundamental theory take on other physical meanings related to geometry and matter. In  \cite{Lin:2017uzr}, Lin proposed that such a separation of entanglement may also be understood in a more general setting, where the ``fundamental'' theory, whose Hilbert space factorizes and does not have a gauge symmetry, has an emergent gauge theory in the IR.\footnote{A gauge theory is emergent if the low energy behavior of the fundamental theory can be identified with that of a gauge theory. We refer the readers to the original reference for the precise definition used in the derivations.} As such, the entanglement entropy $S_{\rm fund}$ of a subregion in the fundamental theory can be written in a form $S_{\rm fund}= S_{\rm edge}+S_{\mathrm{IR}}$. Here $S_{\rm edge}$, which depends on the UV regulator such as a lattice cutoff, takes on the meaning of the analogous area-law term in RT. The IR entropy $S_{\rm IR}$ corresponds to the entanglement of the emergent gauge theory\footnote{Similar ideas appear in the study of emergent gravity in condensed matter models  \cite{Pretko:2017fbf} with emergent gauge theories  \cite{Gu:2009jh}.}. 

Therefore, we may also speculate that a geometry other than AdS emerges from a fundamental theory that is amorphous, in the sense that there are no pre-determined geometric elements. In this case, a generalization of the RT formula (\ref{eqn:RTwCorr}) should still provide a natural separation between UV and IR and identification of the geometric and matter parts of the entanglement without something like a $1/N$ expansion. 
It's worth investigating the prospect that this can be done directly from the state and its associated Hilbert space. 

Here we point out another possible construction proposed by Harlow \cite{Harlow:2016vwg} making use of quantum error correction codes (QECC), or more specifically, the erasure correction codes. A similar RT-like formula is derived in the context of quantum error correction, without having to rely on a background geometry. For the sake of clarity we briefly review some  findings of \cite{Harlow:2016vwg}.

A typical way to protect states against quantum errors is to encode the information non-locally, such that local errors will not easily contaminate the protected information. For instance, let  $|\phi\rangle\in \mathcal{H}_\phi$ be a qudit worth of quantum information. 
One can encode it in a larger Hilbert space
\begin{equation}
\mathcal{H} = (\mathcal{H}_\phi)^{\otimes N}.
\end{equation}
A basis for $\mathcal{H}$ can be formed from the tensor product of basis vectors $|i_j\rangle$ of each copy $j$ of $\mathcal{H}_\phi$. To encode the original state
\begin{equation}
|\phi\rangle = \sum_i C_i |i\rangle \in \mathcal{H}_\phi
\end{equation}
by mapping it to $\mathcal{H}$, we first map each basis element by some fixed rule,
\begin{equation}
  |i\rangle \rightarrow |\tilde i\rangle = \sum_{i_1, i_2 \dots i_N} \mu^{\tilde i}_{i_1i_2\dots i_N}|i_1, i_2 \dots i_N \rangle
\end{equation}
for some coefficients $\mu^{\tilde i}_{i_1i_2\dots i_N}$.
The encoding then takes the form
\begin{equation}
  |\phi\rangle \rightarrow  |\tilde{\phi}\rangle = \sum_{\tilde i} C_{\tilde i}|\tilde i \rangle\in\mathcal{H}.
\end{equation}
The vector subspace of $\mathcal{H}$ spanned by $\{|\tilde i\rangle\}$ is the code subspace, $\mathcal{H}_{\mathrm{code}}$.

To be consistent with the notation in the literature, we will refer to the $N$ qudits making up $\mathcal{H}$ as the physical qudits. 
Let $A$ be a subsystem consisting of a subset of the physical qudits, and $\bar{A}$ its complement, so that 
\begin{equation}
\mathcal{H}=\mathcal{H}_{A}\otimes\mathcal{H}_{\bar{A}}.  
\end{equation}
The encoded information is said to be protected against erasure on $\bar{A}$ if for all 
\begin{equation}
|\tilde{\phi}\rangle\in \mathcal{H}_{\rm code}\subset\mathcal{H},
\end{equation} 
there exists an operator $U_{A}\otimes I_{\bar{A}}$ such that 
\begin{equation}
 U_{A}\otimes I_{\bar{A}}|\tilde{\phi}\rangle_{A\cup\bar{A}} = |\phi\rangle_{j\in A} \otimes|\chi\rangle_{A\cup\bar{A}\setminus\{j\}}
\end{equation}
for some state $|\chi\rangle \in \otimes_{i\neq j}\mathcal{H_\phi}$.
Intuitively, this property allows one to recover the encoded quantum information even though degrees of freedom in $\mathcal{H}_{\bar{A}}$ are inaccessible.

Now consider the scenario in which the code subspace can encode many qudits. One construction is to consider a QECC in which the code subspace factorizes
\begin{equation}
\mathcal{H}_{\rm code}=\mathcal{H}_a\otimes\mathcal{H}_{\bar{a}}.
\end{equation} 
We want the subset $a$ of these code-subspace qudits to be recoverable from the subsystem $A$ of the larger Hilbert space, and similarly the complementary set $\bar{a}$ to be recoverable from $\bar{A}$. Assume each of $\mathcal{H}_{A}, \mathcal{H}_{\bar{A}}$ is further factorizable: 
\begin{equation}
\mathcal{H}_A=\mathcal{H}_{A_1}\otimes \mathcal{H}_{A_2}, \quad \mathcal{H}_{\bar{A}}=\mathcal{H}_{\bar{A_1}}\otimes \mathcal{H}_{\bar{A_2}}, 
\end{equation}
where $\dim \mathcal{H}_{A_1} = \dim\mathcal{H}_{\bar{A_1}} = \dim \mathcal{H}_{\rm code}$. 
As demonstrated by Harlow, a QECC that performs the desired complementary recovery, which satisfies  
\begin{equation}
 |\tilde{i}\rangle| \tilde{j} \rangle = U_{A} U_{\bar{A}}(|i\rangle_{A_1}|j\rangle_{\bar{A_1}}|\chi\rangle_{A_2\bar{A_2}})
\end{equation}
for some entangled state $|\chi\rangle_{A_2\bar{A}_2}$ and unitaries $U_A\otimes \mathbb{I}_{\bar{A}}, \mathbb{I}_{A}\otimes U_{\bar{A}}$, will also satisfy an analogous RT relation. 
Here, $|\tilde{i}\rangle, |\tilde{j}\rangle$,$|i\rangle_{A_1}, |j\rangle_{\bar{A}_1}$ are orthonormal basis vectors for the Hilbert spaces $\mathcal{H}_{a}$, $\mathcal{H}_{\bar{a}}$, $\mathcal{H}_{A_1}$ and $\mathcal{H}_{\bar{A}_1}$ respectively. 

Given a density operator in the code subspace
\begin{equation}
\tilde{\rho}=|\tilde{\phi}\rangle\langle\tilde{\phi}|\in L(\mathcal{H}_{\rm code})\subset L(\mathcal{H}),
\end{equation}
define  reduced density matrices
\begin{align}
&\tilde{\rho}_A = \Tr_{\bar{A}}\tilde{\rho}\\
&\tilde{\rho}_a = \Tr_{\bar{a}}\tilde{\rho}\\
&\rho_{\chi} = \Tr_{\bar{A}_2}|\chi\rangle\langle\chi|.
\end{align}
The resulting RT-like relation for the entropies then takes the form,
\begin{align}
 S(\tilde{\rho}_{A}) &= S(\rho_{\chi})+S(\tilde{\rho}_a)\cr
 S(\tilde{\rho}_{\bar{A}}) &= S(\rho_{\chi})+S(\tilde{\rho}_{\bar{a}}),
\label{eqn:QECCRT}
\end{align}

In the familiar examples of holographic tensor networks and quantum error correction codes \cite{Pastawski:2015qua,Hayden:2016cfa}, the term $S(\rho_{\chi})$ is proportional to the area of the minimal RT surface anchored at the boundary of $A$.\footnote{For example, the entropy $S(\chi)$ is computed by a distillation process and counting the Bell pairs in Sec 4 of  \cite{Pastawski:2015qua}. The cut along the Bell pairs is precisely the bulk minimal surface $\gamma_{A\bar{A}}$ anchored at the boundary of $A$ and $\bar{A}$.}
Consequently, the term $S(\rho_\chi)$ can be understood as the geometric entanglement contribution, while $S(\tilde{\rho}_{a})$ is naturally identified with the ``matter'' contribution to bulk entropy. The sum of these two quantities is equal to $S(\tilde{\rho}_A)$, which is the entanglement entropy of the boundary subregion $A$. 

This generalized RT-like formula can be compared to our equation for the entanglement entropy of a subsystem (\ref{eqn:entropysum}) as the sum of an RC contribution and a subdominant correction.
In particular, the emergent entanglement equilibrium relation (\ref{eqn:EEEC}) can be thought of as the first-order variation of this formula, with the first term representing an area and the second the contribution from the emergent EFT.
In this sense, the entropy formulae underlying BEG can be found more generally in the context of quantum error-correcting codes.
This relation helps shed light on the decomposition of the entanglement entropy into UV geometric contributions and IR contributions from matter fields.

The derivation leading to (\ref{eqn:QECCRT}) makes no reference to a pre-existing geometry or holography. Indeed, it is reasonable to expect such properties to apply to contexts more general than AdS/CFT.  
In particular, conventional QECC as well as the operator-algebra quantum error correction seem directly applicable to bulk entanglement gravity. In fact, they can be used to reconstruct a geometry as long as a notion of entanglement entropy can be consistently defined and computed.  

In the case we are currently interested in, the overall finite-dimensional Hilbert spaces $\mathcal{H}$ in BEG can be identified with the physical Hilbert space in QECC. 
A state $|\psi\rangle$ that encodes geometric information corresponds to the code state $|\tilde{\phi}\rangle\in\mathcal{H}$ above. 
The subsystem $A$ in the form of physical qudits can be identified with some collection of Hilbert space factors (graph vertices) in BEG. 
In addition, the code comes equipped with a code subspace $\mathcal{H}_{\rm code}\subset \mathcal{H}$ which is now identified with the IR subspace of the emergent matter fields. Thus, a natural separation of UV (geometric) and IR (matter) degrees of freedom is simply provided by the subspace or subalgebra associated with a QECC.

Hence, a spatial geometry can be obtained and assigned to quantum error correction codes that do not presume a geometrical interpretation \emph{a priori}. BEG can be particularly useful in the case when one considers deviations from maximally symmetric spaces. For a dynamical theory that preserves the code subspace and Lorentz invariance, the linearized Einstein's equation may emerge as a more generic property, rather than coming from a special theory.

\section{Discussion}

In this work, we have extended the ``space from Hilbert space'' program of \cite{Cao:2016mst}, which posits that spatial geometry can emerge from the entanglement features of appropriate quantum states, to consider the gravitational dynamics of spacetime.
By suggesting a modified entanglement equilibrium condition as well as other assumptions, we are able to sketch how the spacetime metric can be reconstructed from entanglement using the Radon transform, and it how it naturally obeys Einstein's equation at the linearized level.
Our analysis was carried out entirely in (what emerges as) the bulk of spacetime; the entanglement we consider is between Hilbert-space factors representing local degrees of freedom, without reference to AdS/CFT or any other holographic boundary construction.
Colloquially, this bulk entanglement gravity approach can be thought of as finding gravity within quantum mechanics, as opposed to the more conventional approach of quantizing a particular model of spacetime structure.
It also seems to indicate the plausibility of discovering gravitational features from more generic complex quantum systems.

Further work will clearly be required to flesh out this program and put the necessary assumptions on a firmer footing.
We can list a few of the biggest looming questions.
\begin{itemize}
	\item One is to explore the feasibility of developing a specific theory of quantum gravity using quantum information beyond the context of AdS/CFT, for example by specifying an explicit Hamiltonian, but perhaps by less direct means. For instance, a set of constraints on the quantum dynamics could be derived by requiring the emergence of classical general relativity. 
	
	\item Geometry from entanglement is an interesting program all by itself, even without the emergence of gravity. It is important to understand how and if more general emergent geometries, possibly along with their metric tensors, can be reconstructed from entanglement data. 
	
	\item It is also important to address the hope that Lorentz symmetry can be emergent. While there have been discussions mostly in the loop quantum gravity and condensed matter communities, a clear understanding of its feasibility is still lacking. 
	
	\item Given the recent interest in emergent gauge theories in condensed matter models, it may be possible to understand if certain condensed matter models can be ``gravitized'' by emerging the geometry through entanglement of a state, instead of using the pre-existing geometry provided by the lattice structure or Hamiltonian. This may yield interesting toy models that exhibit (analogous) features of gravity.
	
	\item It would be useful to contemplate the emergence of holography from this perspective, going beyond the weak-field gravity context considered here.
\end{itemize}
In addition, one can point to a few more circumscribed and well-defined challenges.
\begin{itemize}
	\item The BEG framework is natural for assigning emergent geometries to tensor networks directly from entanglement. It is also useful for deriving emergent geometry for conventional QECCs as well as their generalizations in the form of operator-algebras. It may be interesting to construct toy models using these concrete tool sets to improve our intuition for the program.
	
	\item Generalizing the tensor Radon transform approach to other Riemannian backgrounds. One particular direction is to make contact with AdS/CFT by considering asymptotically hyperbolic spaces. 
	
	\item Another task is to further understand the UV/IR separation. Since QECC provides a natural separation and a concrete testing ground, specific toy models may be constructed that have non-trivial dynamical properties \cite{Osborne:2017woa}. Efforts in this direction would also improve our understanding in adding backreaction and incorporating general geometries in a tensor network model. On the other hand, geometric characterizations may also help categorize entanglement and code properties. 
	
	\item It would be useful to understand how general the MEEC is in quantum systems near equilibrium, and what physical interpretation can be attached to the two terms.
\end{itemize}

This work has been guided by the conviction that quantum mechanics is the most fundamental theory we have, and implicitly by the Everettian formulation of the theory (the wave function is the only physical variable, and it evolves smoothly and deterministically over time).
In that context, one can argue informally that quantum gravity \emph{must} emerge in roughly the way outlined here.
We human beings generally construct quantum theories by starting with classical theories and quantizing them, but presumably nature doesn't work that way.
There simply is a quantum state, represented by a vector in Hilbert space, evolving according to the Schr\"odinger equation with some particular Hamiltonian. (For these purposes we take time as fundamental, but it is also conceivable that time itself is emergent, arising through the entanglement of ``system'' and ``clock'' factors of Hilbert space.)
Familiar classical concepts such as ``locations in space'' and ``fields'' are necessarily emergent from this basic structure.
Here we have sketched how space and its geometry may plausibly emerge from the entanglement between discrete Hilbert-space factors, and how gravitational dynamics obeying Einstein's equation can be related to entanglement equilibrium.
The is a promising route to a perspective on quantum gravity that puts ``quantum'' first.

\section*{Acknowledgements}
We thank Bartek Czech for the helpful comments on integral geometry that helped inspire this work. We also thank Aidan Chatwin-Davies, Junyu Liu, Spiros Michalakis, and Gunther Uhlmann for engaging and informing discussions. We thank the organizers of YITP long term workshop and QIQG3 at UBC Vancouver. This work is  supported by the U.S. Department of Energy, Office of Science, Office of High Energy Physics, under Award Number DE-SC0011632, as well as by the Walter Burke Institute for Theoretical Physics at Caltech.

\bibliographystyle{JHEP}
\bibliography{Radon}

\providecommand{\href}[2]{#2}\begingroup\raggedright\begin{thebibliography}{10}

\bibitem{Swingle:2009bg}
B.~Swingle, {\it {Entanglement Renormalization and Holography}},  {\em Phys.
  Rev.} {\bf D86} (2012) 065007, [\href{http://arxiv.org/abs/0905.1317}{{\tt
  arXiv:0905.1317}}].

\bibitem{VanRaamsdonk:2010pw}
M.~Van~Raamsdonk, {\it {Building up spacetime with quantum entanglement}},
  {\em Gen. Rel. Grav.} {\bf 42} (2010) 2323--2329,
  [\href{http://arxiv.org/abs/1005.3035}{{\tt arXiv:1005.3035}}]. [Int. J. Mod.
  Phys.D19,2429(2010)].

\bibitem{TNSGeo}
G.~{Evenbly} and G.~{Vidal}, {\it {Tensor Network States and Geometry}},  {\em
  Journal of Statistical Physics} {\bf 145} (Nov., 2011) 891--918,
  [\href{http://arxiv.org/abs/1106.1082}{{\tt arXiv:1106.1082}}].

\bibitem{Faulkner:2013ica}
T.~Faulkner, M.~Guica, T.~Hartman, R.~C. Myers, and M.~Van~Raamsdonk, {\it
  {Gravitation from Entanglement in Holographic CFTs}},  {\em JHEP} {\bf 03}
  (2014) 051, [\href{http://arxiv.org/abs/1312.7856}{{\tt arXiv:1312.7856}}].

\bibitem{Faulkner:2017tkh}
T.~Faulkner, F.~M. Haehl, E.~Hijano, O.~Parrikar, C.~Rabideau, and
  M.~Van~Raamsdonk, {\it {Nonlinear Gravity from Entanglement in Conformal
  Field Theories}},  {\em JHEP} {\bf 08} (2017) 057,
  [\href{http://arxiv.org/abs/1705.03026}{{\tt arXiv:1705.03026}}].

\bibitem{Czech:2015kbp}
B.~Czech, L.~Lamprou, S.~McCandlish, and J.~Sully, {\it {Tensor Networks from
  Kinematic Space}},  {\em JHEP} {\bf 07} (2016) 100,
  [\href{http://arxiv.org/abs/1512.01548}{{\tt arXiv:1512.01548}}].

\bibitem{Maldacena:1997re}
J.~M. Maldacena, {\it {The Large N limit of superconformal field theories and
  supergravity}},  {\em Int. J. Theor. Phys.} {\bf 38} (1999) 1113--1133,
  [\href{http://arxiv.org/abs/hep-th/9711200}{{\tt hep-th/9711200}}]. [Adv.
  Theor. Math. Phys.2,231(1998)].

\bibitem{Swingle:2014uza}
B.~Swingle and M.~Van~Raamsdonk, {\it {Universality of Gravity from
  Entanglement}},  \href{http://arxiv.org/abs/1405.2933}{{\tt
  arXiv:1405.2933}}.

\bibitem{Maldacena:2013xja}
J.~Maldacena and L.~Susskind, {\it {Cool horizons for entangled black holes}},
  {\em Fortsch. Phys.} {\bf 61} (2013) 781--811,
  [\href{http://arxiv.org/abs/1306.0533}{{\tt arXiv:1306.0533}}].

\bibitem{Sanches:2016sxy}
F.~Sanches and S.~J. Weinberg, {\it {Holographic entanglement entropy
  conjecture for general spacetimes}},  {\em Phys. Rev.} {\bf D94} (2016),
  no.~8 084034, [\href{http://arxiv.org/abs/1603.05250}{{\tt
  arXiv:1603.05250}}].

\bibitem{Susskind:2014moa}
L.~Susskind, {\it {Entanglement is not enough}},  {\em Fortsch. Phys.} {\bf 64}
  (2016) 49--71, [\href{http://arxiv.org/abs/1411.0690}{{\tt
  arXiv:1411.0690}}].

\bibitem{Cao:2016mst}
C.~Cao, S.~M. Carroll, and S.~Michalakis, {\it {Space from Hilbert Space:
  Recovering Geometry from Bulk Entanglement}},  {\em Phys. Rev.} {\bf D95}
  (2017), no.~2 024031, [\href{http://arxiv.org/abs/1606.08444}{{\tt
  arXiv:1606.08444}}].

\bibitem{Bao:2017iye}
N.~Bao, C.~Cao, S.~M. Carroll, and L.~McAllister, {\it {Quantum Circuit
  Cosmology: The Expansion of the Universe Since the First Qubit}},
  \href{http://arxiv.org/abs/1702.06959}{{\tt arXiv:1702.06959}}.

\bibitem{Bao:2017qmt}
N.~Bao, C.~Cao, S.~M. Carroll, and A.~Chatwin-Davies, {\it {De Sitter Space as
  a Tensor Network: Cosmic No-Hair, Complementarity, and Complexity}},
  \href{http://arxiv.org/abs/1709.03513}{{\tt arXiv:1709.03513}}.

\bibitem{Susskind:2017ney}
L.~Susskind, {\it {Dear Qubitzers, GR=QM}},
  \href{http://arxiv.org/abs/1708.03040}{{\tt arXiv:1708.03040}}.

\bibitem{Jacobson:1995ab}
T.~Jacobson, {\it {Thermodynamics of space-time: The Einstein equation of
  state}},  {\em Phys. Rev. Lett.} {\bf 75} (1995) 1260--1263,
  [\href{http://arxiv.org/abs/gr-qc/9504004}{{\tt gr-qc/9504004}}].

\bibitem{Padmanabhan:2009vy}
T.~Padmanabhan, {\it {Thermodynamical Aspects of Gravity: New insights}},  {\em
  Rept. Prog. Phys.} {\bf 73} (2010) 046901,
  [\href{http://arxiv.org/abs/0911.5004}{{\tt arXiv:0911.5004}}].

\bibitem{Verlinde:2010hp}
E.~P. Verlinde, {\it {On the Origin of Gravity and the Laws of Newton}},  {\em
  JHEP} {\bf 04} (2011) 029, [\href{http://arxiv.org/abs/1001.0785}{{\tt
  arXiv:1001.0785}}].

\bibitem{Jacobson:2015hqa}
T.~Jacobson, {\it {Entanglement Equilibrium and the Einstein Equation}},  {\em
  Phys. Rev. Lett.} {\bf 116} (2016), no.~20 201101,
  [\href{http://arxiv.org/abs/1505.04753}{{\tt arXiv:1505.04753}}].

\bibitem{Carroll:2016lku}
S.~M. Carroll and G.~N. Remmen, {\it {What is the Entropy in Entropic
  Gravity?}},  {\em Phys. Rev.} {\bf D93} (2016), no.~12 124052,
  [\href{http://arxiv.org/abs/1601.07558}{{\tt arXiv:1601.07558}}].

\bibitem{Verlinde:2016toy}
E.~P. Verlinde, {\it {Emergent Gravity and the Dark Universe}},  {\em SciPost
  Phys.} {\bf 2} (2017), no.~3 016,
  [\href{http://arxiv.org/abs/1611.02269}{{\tt arXiv:1611.02269}}].

\bibitem{Banks:2010tj}
T.~Banks, {\it {TASI Lectures on Holographic Space-Time, SUSY and Gravitational
  Effective Field Theory}},  in {\em {Proceedings, Theoretical Advanced Study
  Institute in Elementary Particle Physics (TASI 2010). String Theory and Its
  Applications: From meV to the Planck Scale: Boulder, Colorado, USA, June
  1-25, 2010}}, 2010.
\newblock \href{http://arxiv.org/abs/1007.4001}{{\tt arXiv:1007.4001}}.

\bibitem{Banks:2011av}
T.~Banks, {\it {Holographic Space-Time: The Takeaway}},
  \href{http://arxiv.org/abs/1109.2435}{{\tt arXiv:1109.2435}}.

\bibitem{Banks:2015iya}
T.~Banks and W.~Fischler, {\it {Holographic Inflation Revised}},
  \href{http://arxiv.org/abs/1501.01686}{{\tt arXiv:1501.01686}}.

\bibitem{Nomura:2016aww}
Y.~Nomura, N.~Salzetta, F.~Sanches, and S.~J. Weinberg, {\it {Spacetime Equals
  Entanglement}},  {\em Phys. Lett.} {\bf B763} (2016) 370--374,
  [\href{http://arxiv.org/abs/1607.02508}{{\tt arXiv:1607.02508}}].

\bibitem{Nomura:2016ikr}
Y.~Nomura, N.~Salzetta, F.~Sanches, and S.~J. Weinberg, {\it {Toward a
  Holographic Theory for General Spacetimes}},  {\em Phys. Rev.} {\bf D95}
  (2017), no.~8 086002, [\href{http://arxiv.org/abs/1611.02702}{{\tt
  arXiv:1611.02702}}].

\bibitem{sharafutdinov1994integral}
V.~Sharafutdinov, {\em Integral Geometry of Tensor Fields}.
\newblock Inverse and ill-posed problems series. VSP, 1994.

\bibitem{Harlow:2016vwg}
D.~Harlow, {\it {The Ryu–Takayanagi Formula from Quantum Error Correction}},
  {\em Commun. Math. Phys.} {\bf 354} (2017), no.~3 865--912,
  [\href{http://arxiv.org/abs/1607.03901}{{\tt arXiv:1607.03901}}].

\bibitem{Czech:2016tqr}
B.~Czech, L.~Lamprou, S.~McCandlish, B.~Mosk, and J.~Sully, {\it {Equivalent
  Equations of Motion for Gravity and Entropy}},  {\em JHEP} {\bf 02} (2017)
  004, [\href{http://arxiv.org/abs/1608.06282}{{\tt arXiv:1608.06282}}].

\bibitem{Czech:2015qta}
B.~Czech, L.~Lamprou, S.~McCandlish, and J.~Sully, {\it {Integral Geometry and
  Holography}},  {\em JHEP} {\bf 10} (2015) 175,
  [\href{http://arxiv.org/abs/1505.05515}{{\tt arXiv:1505.05515}}].

\bibitem{Cotler:2017abq}
J.~S. Cotler, G.~R. Penington, and D.~H. Ranard, {\it {Locality from the
  Spectrum}},  \href{http://arxiv.org/abs/1702.06142}{{\tt arXiv:1702.06142}}.

\bibitem{Casini:2008cr}
H.~Casini, {\it {Relative entropy and the Bekenstein bound}},  {\em Class.
  Quant. Grav.} {\bf 25} (2008) 205021,
  [\href{http://arxiv.org/abs/0804.2182}{{\tt arXiv:0804.2182}}].

\bibitem{Bousso:2014sda}
R.~Bousso, H.~Casini, Z.~Fisher, and J.~Maldacena, {\it {Proof of a Quantum
  Bousso Bound}},  {\em Phys. Rev.} {\bf D90} (2014), no.~4 044002,
  [\href{http://arxiv.org/abs/1404.5635}{{\tt arXiv:1404.5635}}].

\bibitem{Bousso:2014uxa}
R.~Bousso, H.~Casini, Z.~Fisher, and J.~Maldacena, {\it {Entropy on a null
  surface for interacting quantum field theories and the Bousso bound}},  {\em
  Phys. Rev.} {\bf D91} (2015), no.~8 084030,
  [\href{http://arxiv.org/abs/1406.4545}{{\tt arXiv:1406.4545}}].

\bibitem{Bao:2017rnv}
N.~Bao, S.~M. Carroll, and A.~Singh, {\it {The Hilbert Space of Quantum Gravity
  Is Locally Finite-Dimensional}},  \href{http://arxiv.org/abs/1704.00066}{{\tt
  arXiv:1704.00066}}.

\bibitem{Bao:2015boa}
N.~Bao, C.~Cao, M.~Walter, and Z.~Wang, {\it {Holographic entropy inequalities
  and gapped phases of matter}},  {\em JHEP} {\bf 09} (2015) 203,
  [\href{http://arxiv.org/abs/1507.05650}{{\tt arXiv:1507.05650}}].

\bibitem{AreaLawEntRev}
J.~{Eisert}, M.~{Cramer}, and M.~B. {Plenio}, {\it {Colloquium: Area laws for
  the entanglement entropy}},  {\em Reviews of Modern Physics} {\bf 82} (Jan.,
  2010) 277--306, [\href{http://arxiv.org/abs/0808.3773}{{\tt
  arXiv:0808.3773}}].

\bibitem{Orus:2013kga}
R.~Orus, {\it {A Practical Introduction to Tensor Networks: Matrix Product
  States and Projected Entangled Pair States}},  {\em Annals Phys.} {\bf 349}
  (2014) 117--158, [\href{http://arxiv.org/abs/1306.2164}{{\tt
  arXiv:1306.2164}}].

\bibitem{Pastawski:2015qua}
F.~Pastawski, B.~Yoshida, D.~Harlow, and J.~Preskill, {\it {Holographic quantum
  error-correcting codes: Toy models for the bulk/boundary correspondence}},
  {\em JHEP} {\bf 06} (2015) 149, [\href{http://arxiv.org/abs/1503.06237}{{\tt
  arXiv:1503.06237}}].

\bibitem{Hayden:2016cfa}
P.~Hayden, S.~Nezami, X.-L. Qi, N.~Thomas, M.~Walter, and Z.~Yang, {\it
  {Holographic duality from random tensor networks}},  {\em JHEP} {\bf 11}
  (2016) 009, [\href{http://arxiv.org/abs/1601.01694}{{\tt arXiv:1601.01694}}].

\bibitem{Bianchi:2012ev}
E.~Bianchi and R.~C. Myers, {\it {On the Architecture of Spacetime Geometry}},
  {\em Class. Quant. Grav.} {\bf 31} (2014) 214002,
  [\href{http://arxiv.org/abs/1212.5183}{{\tt arXiv:1212.5183}}].

\bibitem{Myers:2013lva}
R.~C. Myers, R.~Pourhasan, and M.~Smolkin, {\it {On Spacetime Entanglement}},
  {\em JHEP} {\bf 06} (2013) 013, [\href{http://arxiv.org/abs/1304.2030}{{\tt
  arXiv:1304.2030}}].

\bibitem{cmds}
J.~{Kruscal} and M.~{Wish}, {\it {Multidimensional Scaling}},  {\em Sage
  University Paper series on Quantitative Application in the Social Sciences}
  {\bf 07} (1978) 011.

\bibitem{walter}
J.~Walter and H.~Ritter, {\it On interactive visualization of high-dimensional
  data using the hyperbolic plane},  in {\em Proceedings of the Eighth ACM
  SIGKDD International Conference on Knowledge Discovery and Data Mining}, KDD
  '02, (New York, NY, USA), pp.~123--132, ACM, 2002.

\bibitem{eladkimmel}
A.~Elad and R.~Kimmel, {\it Spherical flattening of the cortex surface},  in
  {\em Geometric Methods in Bio-Medical Image Processing} (R.~Malladi, ed.),
  Mathematics and Visualization, pp.~77--89.
\newblock Springer Berlin Heidelberg, 2002.

\bibitem{gmds}
A.~M. Bronstein, M.~M. Bronstein, and R.~Kimmel, {\it {Generalized
  multidimensional scaling: A framework for isometry-invariant partial surface
  matching}},  {\em Proceedings of the National Academy of Science} (2006)
  1168--1172.

\bibitem{mdsbook}
I.~Borg and P.~Groenen, {\em {Modern Multidimensional Scaling: Theory and
  Applications}}.
\newblock Springer, 2005.

\bibitem{BRP2dproof}
L.~{Pestov} and G.~{Uhlmann}, {\it {Two dimensional compact simple Riemannian
  manifolds are boundary distance rigid}},  {\em ArXiv Mathematics e-prints}
  (May, 2003) [\href{http://arxiv.org/abs/math/0305280}{{\tt math/0305280}}].

\bibitem{uhlmann}
G.~Uhlmann in prep.

\bibitem{helgason1999Radon}
S.~Helgason, {\em The Radon Transform}.
\newblock Progress in Mathematics. Birkh{\"a}user Boston, 1999.

\bibitem{BRPrev1}
P.~Stefanov and G.~Uhlmann, {\it Recent progress on the boundary rigidity
  problem},  {\em Electron. Res. Announc. Amer. Math. Soc.} {\bf 11} (Jun,
  2005) 64--70.

\bibitem{Stefanov2008}
P.~Stefanov and G.~Uhlmann, {\em Boundary and lens rigidity, tensor tomography
  and analytic microlocal analysis}, pp.~275--293.
\newblock Springer Japan, Tokyo, 2008.

\bibitem{john1938}
F.~John, {\it The ultrahyperbolic differential equation with four independent
  variables},  {\em Duke Math. J.} {\bf 4} (06, 1938) 300--322.

\bibitem{michel}
R.~Michel, {\it Sur la rigidit\'e impos\'ee par la longueur des
  g\'eod\'esiques},  {\em Invent. Math.} {\bf 65} (1981) 71--83.

\bibitem{BRPflat}
D.~Burago and S.~Ivanov, {\it Boundary rigidity and filling volume minimality
  of metrics close to a flat one},  {\em Annals of Mathematics} {\bf 171}
  (2010), no.~2 1183--1211.

\bibitem{BRPrev2}
G.~P. {Paternain}, M.~{Salo}, and G.~{Uhlmann}, {\it {Tensor tomography:
  progress and challenges}},  {\em ArXiv e-prints} (Mar., 2013)
  [\href{http://arxiv.org/abs/1303.6114}{{\tt arXiv:1303.6114}}].

\bibitem{Croke2004}
C.~B. Croke, {\em Rigidity Theorems in Riemannian Geometry}, pp.~47--72.
\newblock Springer New York, New York, NY, 2004.

\bibitem{Sharafutdinov2007}
V.~Sharafutdinov, {\it Variations of dirichlet-to-neumann map and deformation
  boundary rigidity of simple 2-manifolds},  {\em The Journal of Geometric
  Analysis} {\bf 17} (Mar, 2007) 147.

\bibitem{Stefanov2014}
P.~{Stefanov}, G.~{Uhlmann}, and A.~{Vasy}, {\it {Inverting the local geodesic
  X-ray transform on tensors}},  {\em ArXiv e-prints} (Oct., 2014)
  [\href{http://arxiv.org/abs/1410.5145}{{\tt arXiv:1410.5145}}].

\bibitem{Monard}
F.~Monard, {\it Numerical implementation of geodesic x-ray transforms and their
  inversion},  {\em SIAM Journal on Imaging Sciences} {\bf 7} (2014), no.~2
  1335--1357,
  [\href{http://arxiv.org/abs/https://doi.org/10.1137/130938657}{{\tt
  https://doi.org/10.1137/130938657}}].

\bibitem{ising1}
R.~J. Elliott and C.~Wood, {\it The ising model with a transverse field. i.
  high temperature expansion},  {\em Journal of Physics C: Solid State Physics}
  {\bf 4} (1971), no.~15 2359.

\bibitem{ising2}
P.~Pfeuty and R.~J. Elliott, {\it The ising model with a transverse field. ii.
  ground state properties},  {\em Journal of Physics C: Solid State Physics}
  {\bf 4} (1971), no.~15 2370.

\bibitem{Kitaev97}
A.~Y. {Kitaev}, {\it {Fault-tolerant quantum computation by anyons}},  {\em
  Annals of Physics} {\bf 303} (Jan., 2003) 2--30,
  [\href{http://arxiv.org/abs/quant-ph/9707021}{{\tt quant-ph/9707021}}].

\bibitem{1165108}
G.~Beylkin, {\it Discrete radon transform},  {\em IEEE Transactions on
  Acoustics, Speech, and Signal Processing} {\bf 35} (Feb, 1987) 162--172.

\bibitem{finitequotient}
C.~Croke, {\it Boundary and lens rigidity of finite quotients},  {\em
  Proceedings of the American Mathematical Society} {\bf 133} (2005), no.~12
  3663--3668.

\bibitem{torus2}
C.~B. Croke, {\it Volumes of balls in manifolds without conjugate points},
  {\em International Journal of Mathematics} {\bf 03} (1992), no.~04 455--467,
  [\href{http://arxiv.org/abs/http://www.worldscientific.com/doi/pdf/10.1142/S0129167X92000205}{{\tt
  http://www.worldscientific.com/doi/pdf/10.1142/S0129167X92000205}}].

\bibitem{croke1991}
C.~B. Croke, {\it Rigidity and the distance between boundary points},  {\em J.
  Differential Geom.} {\bf 33} (1991), no.~2 445--464.

\bibitem{Tracey2017}
S.~{Alexakis}, T.~{Balehowsky}, and A.~{Nachman}, {\it {Determining a
  Riemannian Metric from Minimal Areas}},  {\em ArXiv e-prints} (Nov., 2017)
  [\href{http://arxiv.org/abs/1711.09379}{{\tt arXiv:1711.09379}}].

\bibitem{Porrati:2003na}
M.~Porrati and R.~Rabadan, {\it {Boundary rigidity and holography}},  {\em
  JHEP} {\bf 01} (2004) 034, [\href{http://arxiv.org/abs/hep-th/0312039}{{\tt
  hep-th/0312039}}].

\bibitem{Gourgoulhon:2007ue}
E.~Gourgoulhon, {\it {3+1 formalism and bases of numerical relativity}},
  \href{http://arxiv.org/abs/gr-qc/0703035}{{\tt gr-qc/0703035}}.

\bibitem{Blanco:2013joa}
D.~D. Blanco, H.~Casini, L.-Y. Hung, and R.~C. Myers, {\it {Relative Entropy
  and Holography}},  {\em JHEP} {\bf 08} (2013) 060,
  [\href{http://arxiv.org/abs/1305.3182}{{\tt arXiv:1305.3182}}].

\bibitem{Faulkner:2013ana}
T.~Faulkner, A.~Lewkowycz, and J.~Maldacena, {\it {Quantum corrections to
  holographic entanglement entropy}},  {\em JHEP} {\bf 11} (2013) 074,
  [\href{http://arxiv.org/abs/1307.2892}{{\tt arXiv:1307.2892}}].

\bibitem{Lin:2017uzr}
J.~Lin, {\it {Ryu-Takayanagi Area as an Entanglement Edge Term}},
  \href{http://arxiv.org/abs/1704.07763}{{\tt arXiv:1704.07763}}.

\bibitem{Pretko:2017fbf}
M.~Pretko, {\it {Emergent gravity of fractons: Mach’s principle revisited}},
  {\em Phys. Rev.} {\bf D96} (2017), no.~2 024051,
  [\href{http://arxiv.org/abs/1702.07613}{{\tt arXiv:1702.07613}}].

\bibitem{Gu:2009jh}
Z.-C. Gu and X.-G. Wen, {\it {Emergence of helicity +- 2 modes (gravitons) from
  qbit models}},  {\em Nucl. Phys.} {\bf B863} (2012) 90--129,
  [\href{http://arxiv.org/abs/0907.1203}{{\tt arXiv:0907.1203}}].

\bibitem{Osborne:2017woa}
T.~J. Osborne and D.~E. Stiegemann, {\it {Dynamics for holographic codes}},
  \href{http://arxiv.org/abs/1706.08823}{{\tt arXiv:1706.08823}}.

\end{thebibliography}\endgroup

\end{document}